\documentclass[12pt]{article}
\usepackage{pic03}
\usepackage{hyperref}
\usepackage{url}
\usepackage{graphicx}

\input{tcilatex}

\begin{document}

\title{\textbf{FUTURE LEPTON HADRON COLLIDERS\ }}
\author{ \"{O}mer Yava\c{s} \\
\emph{Ankara University, Fac. of Engineering, Dept. of Eng. of Physics }\\
\emph{\ 06100 Tandogan, Ankara, TURKEY}}
\maketitle

\begin{abstract}
Main parameters of future lepton-hadron colliders are estimated. Namely,
THERA and Linac*LHC based ep, $\gamma $p, eA, $\gamma $A and FEL$\gamma $A
colliders are considered. The physics search potential of these machines are
considered.
\end{abstract}

\baselineskip=14.5pt

\baselineskip=17pt

\section{Introduction}

It is known that lepton-hadron collisions have been playing a crucial role
in exploration of deep inside of matter. Today, THERA (TESLA on HERA) and
Linac*LHC can be considered as realistic candidates for future lepton-hadron
and photon hadron colliders. We discuss the main parameters and physics
search potential of the lepton-hadron colliders and draw the attention of
the high energy and nuclear physics communities to future $ep,eA,\gamma
p,\gamma A$ and FEL $\gamma A$ collider facilities.

\section{TESLA*HERA Based Lepton-Hadron Colliders}

It is known that the TESLA Collider will be a powerful tool for exploraiton
of the multi-hundred GeV scale \cite{Brinkmann1}. Taking into account the
possible polarized proton and nucleus options for HERA will provide a number
of additional opportunities to investigate lepton-hadron and photon hadron
interactions at TeV scale. Recently, the work on TESLA TDR has been
finished, and TESLA$\otimes $HERA based $ep,\gamma p,eA$ and $\gamma A$
colliders are included into TESLA\ project. Main parameters of TESLA$\otimes 
$HERA based $ep$ collider are given in \cite{therabook}. It is seen that one
has $L_{ep}=4.1\cdot 10^{30}$ $cm^{-2}s^{-1}$\ with $E_{e}=250$ GeV and $%
E_{p}=1$ $TeV$. Also two additional versions ($E_{e}=E_{p}=500$ $GeV$ with $%
L_{ep}=2.5\cdot 10^{31}$ $cm^{-2}s^{-1}$ and $E_{e}=$ $E_{p}=800$ $GeV$ with 
$L_{ep}=1.6\cdot 10^{31}$ $cm^{-2}s^{-1}$) have been mentioned in TESLA TDR.
In principle, TESLA$\otimes $ HERA based ep collider will extend the HERA
kinematics region by an order in both Q$^{2}$ and $x$ and, therefore, the
parton saturation regime can be achieved. Main parameters and physics search
potential of THERA based $\gamma p$\textbf{\ }collider are given in \cite%
{Ciftci95, Brinkmann97}. Main limitations for $eA$ option comes from fast
emittance growth of nucleus beam due to intra-beam scattering. In our
opinion $\gamma A$ option is the most promising option of TESLA$\otimes $%
HERA complex, because it will give unique opportunity to investigate small $%
x_{g}$ region in nuclear medium. Colliding of TESLA FEL beam with nucleus
bunches from HERA may give a unique possibility to investigate
\textquotedblright old\textquotedblright\ nuclear phenomena in rather
unusual conditions. The main idea is very simple \cite{Aktas99}:
ultra-relativistic ions will see laser photons with energy $\omega _{o}$\ as
a beam of photons with energy 2$\gamma _{A}\omega _{o}$\ , where $\gamma
_{A} $\ is the Lorentz factor of the ion beam. The region 0.1$\div $10 MeV,
which is matter of interest for nuclear spectroscopy, corresponds to 0.1$%
\div $10 keV lasers, which coincide with the energy region of TESLA FEL.

\section{Linac*LHC Based Lepton-Hadron Colliders}

The center-of-mass energies which will be achieved at different options of
this machine \cite{Yavas00} are an order larger than those at HERA are and $%
\sim $3 times larger than the energy region of TESLA$\otimes $HERA.
Center-of-mass energy and luminosity for this option are $\sqrt{s}=5.29$ TeV
and $L_{ep}=8\cdot 10^{31}$ cm$^{-2}$s$^{-1}.$ This machine, which will
extend both the $Q^{2}$-range and $\ x$-range by more than two order of
magnitude comparing to those explored by HERA, has a strong potential for
both standard model and new physics research. Using $\gamma p$\textbf{\ }%
option\textbf{\ }of\textbf{\ }this collider thousands di-jets with $%
p_{t}>500 $\ GeV and hundreds thousands single W bosons will be produced,
hundred millions of \ $\overline{b}b$- and \ $\overline{c}c$- pairs will
give opportunity to explore the region of extremely small x$_{g}$ etc.
Details on main parameters and physics search potential of Linac*LHC based $%
eA$, $\gamma A$ and FEL $\gamma A$\ coliders can be found in \cite%
{Brinkmann97,Yavas00}. The CLIC, an electron-positron collider with $\sqrt{s}%
=3$ $TeV$ and $L_{ee}=10^{35}$ $cm^{-2}s^{-1},$ is considered as one of the
future options for post-LHC era at CERN. The work on CLIC*LHC based $%
ep,\gamma p,eA,\gamma A$ and FEL $\gamma A$ options is under progress.

\section{Conclusions}

It seems that neither HERA nor LHC$\otimes $LEP will be the end points for
lepton-hadron colliders. We see that TeV scale linac-ring type $ep$ machines
will give an opportunity to go far in this direction (see Table 1). In
addition, more knowledge on the subject can be found in \cite{Ciftci03}.

Table 1: \textit{Future lepton-hadron colliders: a) First stage (2010-2015).}

{%
\begin{tabular}{|c|c|c|c|}
\hline
& TESLA$\otimes $HERA & LEP$\otimes $LHC & e$\otimes $RHIC \\ \hline
$\sqrt{s},$ TeV & 1.0$\rightarrow $1.6 & 1.37 & 0.1 \\ \hline
$E_{l},$ TeV & 0.25$\rightarrow $0.8 & 0.0673 & 0.01 \\ \hline
$E_{p},$ TeV & 1 & 7 & 0.25 \\ \hline
$L,$ 10$^{31}$ $cm^{-2}s^{-1}$ & 1-10 & 12 & 46 \\ \hline
Additional options & $eA,\gamma p,\gamma A,$ $FEL\gamma A$ & $eA$ & $%
eA,FEL\gamma A$ \\ \hline
\end{tabular}%
}

b) \textit{Second stage (2015-2020)}.

{%
\begin{tabular}{|c|c|c|}
\hline
& Linac$\otimes $LHC & CLIC based \\ \hline
$\sqrt{s},$ TeV & 5.29 & 3 \\ \hline
$E_{l},$ TeV & 1 & 1.5 \\ \hline
$E_{p},$ TeV & 7 & 1.5 \\ \hline
$L,10^{31}$ $cm^{-2}s^{-1}$ & 10-100 & 10 \\ \hline
Additional options & $eA,\gamma p,\gamma A,$ $FEL\gamma A$ & $eA,\gamma
p,\gamma A,$ $FEL\gamma A$ \\ \hline
\end{tabular}%
}

\bigskip

This work is partially supported by Turkish State Planning Organization
under the Grant No 2002 K 120250.

\bigskip

\bigskip

\end{document}